\documentclass[prl,reprint,amsmath,amssymb,aps,preprintnumbers,showpacs]{revtex4-1}
\usepackage[colorlinks=true,linkcolor=black, citecolor=black,urlcolor=black]{hyperref} 
\usepackage{amstext,amsmath,amssymb,amsfonts,mathrsfs}   
\usepackage{graphicx}  
\usepackage{braket} 
\usepackage{bbm} 
\usepackage{tikz} 
\usetikzlibrary{calc}
\usetikzlibrary{arrows}

\newcommand{\AdS}{\text{AdS}}
\renewcommand{\S}{\text{S}}
\newcommand{\T}{\text{T}}
\newcommand{\CFT}{\text{CFT}}

\newcommand{\alg}[1]{\mathfrak{#1}}

\newcommand{\so}{\alg{so}}
\renewcommand{\u}{\alg{u}}
\newcommand{\su}{\alg{su}}
\newcommand{\psu}{\alg{psu}}

\newcommand{\de}{\text{d}}

\newcommand{\suA}{\bullet}
\newcommand{\suB}{\circ}

\newcommand{\ce}{\text{c.e.}}

\newcommand{\stL}{\tilde{\text{\tiny L}}}

\newcommand{\Smat}{\mathbf{S}}

\newcommand{\Ccr}{\mathscr{C}}
\newcommand{\id}{\mathbf{1}}

\newcommand{\sL}{\mbox{\tiny L}}
\newcommand{\sR}{\mbox{\tiny R}}
\newcommand{\bL}{\text{L}}
\newcommand{\bR}{\text{R}}

\newcommand{\gen}[1]{\mathbf{#1}}
\newcommand{\sgen}[1]{\mathbf{#1}}

\newcommand{\acomm}[2]{\{#1,#2\}}

\begin{document}
\preprint{Imperial-TP-OOS-2014-01, HU-Mathematik-2014-05}
\preprint{HU-EP-14/12, ITP-UU-14/10, SPIN-14/11}

\vspace*{0cm}

\title{All-loop worldsheet S~matrix for \texorpdfstring{$\AdS_3\times\S^3\times\T^4$}{AdS(3)xS**3xT**4}}
\author{Riccardo Borsato$^{1}$}
\email{R.Borsato@uu.nl}
\author{Olof Ohlsson Sax$^{2}$}
\email{o.olsson-sax@imperial.ac.uk}
\author{Alessandro Sfondrini$^{1,3}$}
\email{Alessandro.Sfondrini@physik.hu-berlin.de}
\author{Bogdan Stefa{\'n}ski, jr.$^{4}$}
\email{Bogdan.Stefanski.1@city.ac.uk}

\affiliation{$^1$ Institute for Theoretical Physics \& Spinoza Institute, Utrecht University, 3508 TD Utrecht, The Netherlands \\
${}^2$ The Blackett Laboratory, Imperial College, London SW7 2AZ, United Kingdom \\
${}^3$ Institut f{\"u}r Mathematik \& Institut f{\"u}r Physik, Humboldt-Universit{\"a}t zu Berlin, Zum Gro{\ss}en Windkanal 6, 12489 Berlin, Germany \\
${}^4$ Centre for Mathematical Science, City University London, Northampton Square, London EC1V 0HB, United Kingdom}

\begin{abstract}
We obtain the all-loop worldsheet S~matrix for fundamental excitations on $\AdS_3\times\S^3\times\T^4$ by studying the off-shell symmetry algebra of the superspace action in lightcone gauge.
The massless modes, unaccounted for in earlier works, are automatically included in our treatment. Their exact dispersion relation is found to be non-relativistic, of giant-magnon form and their scattering is naturally well-defined.
This opens the way to a complete investigation of~$\AdS_3/\CFT_2$ integrability.
\end{abstract}


\maketitle

\section{Introduction}
Recent years have witnessed remarkable progresses in understanding the gauge/string correspondence using integrability methods; see~\cite{Arutyunov:2009ga,Beisert:2010jr,future:review} for reviews. Given the power of these methods to study non-protected quantities at any value of the coupling,
it is particularly interesting to apply them to the $\AdS_3/\CFT_2$ dual pairs. Such pairs were amongst the first examples of holography~\cite{Brown:1986nw}. They feature the infinite-dimensional Virasoro symmetry algebra, allow for black hole solutions~\cite{Banados:1992wn} and play an important role in string-theoretical black-hole microstate counting~\cite{Strominger:1996sh}.

In string theory, $\AdS_3\times\S^3\times\T^4$ emerges from the near-horizon limit of a system of~D1 and~D5 branes. As a result, the gauge theory has fundamental as well as adjoint matter fields. The dual pair has an infinite-dimensional $\mathcal{N}=(4,4)$ superconformal symmetry~\cite{Maldacena:1997re,Elitzur:1998mm,Larsen:1999uk,Seiberg:1999xz}.
The background preserves relatively little supersymmetry---sixteen supercharges---and can be supported by a mixture of Ramond-Ramond (RR) and Neveu-Schwarz-Neveu-Schwarz (NSNS) fluxes.
While the pure-NSNS background can be studied by worldsheet CFT techniques~\cite{Giveon:1998ns, Maldacena:2000hw}, it is the pure-RR one that naturally emerges from the near-horizon limit of D-branes and so is expected to give the description of the dual gauge theory in the strongly-coupled planar limit. The RR and NSNS descriptions are related by the non-perturbative and non-planar string-theory S~duality. 
Therefore, directly understanding the RR background is an important goal in the study of $\AdS_3/\CFT_2$.

Classical integrability for string theory on $\AdS_3\times\S^3\times\T^4$
was established in Refs.~\cite{Babichenko:2009dk,Sundin:2012gc,Cagnazzo:2012se}. As was observed in Ref.~\cite{Babichenko:2009dk}, the presence of flat directions in this background gives rise to massless string modes, which, at first sight, cannot be analysed using integrability methods. Classically, this was addressed only recently~\cite{Lloyd:2013wza}, while so far quantum integrability of the worldsheet theory was probed only for massive modes~\cite{Borsato:2012ud, Borsato:2012ss, Borsato:2013qpa, Hoare:2013ida, Hoare:2013lja}; see however Ref.~\cite{Sax:2012jv} for a description of massless modes in the weakly-coupled spin chain. Massless modes are notoriously problematic for integrable scattering~\cite{Zamolodchikov:1992zr, Fendley:1993jh}.
In this letter we show how to incorporate massless modes of pure-RR $\AdS_3\times\S^3\times\T^4$ strings into the integrability machinery by presenting the complete all-loop S~matrix for fundamental worldsheet excitations.

In $1+1$~dimensions, quantum integrability manifests itself as factorisation of the S~matrix---any scattering process decomposes into sequences of two-body ones~\cite{Zamolodchikov:1978xm}. Consistency of this decomposition requires a cubic identity for the two-body S~matrix~$\Smat$, the celebrated~\emph{Yang-Baxter equation}. Once~$\Smat$ is known, the energy spectrum of the theory can be found by Bethe ansatz techniques.
Here, the observable is the energy spectrum of closed $\AdS_3\times\S^3\times\T^4$ strings in the planar limit. Therefore we consider the non-linear $\sigma$~model (NLSM) from the $1+1$ dimensional worldsheet into~$\AdS_3\times\S^3\times\T^4$, and compute its worldsheet S~matrix. 

To this end, we first study the symmetries of the lightcone gauge-fixed NLSM, in a limit where the closed-string worldsheet cylinder decompactifies to a plane and~$\Smat$ is well-defined. Unlike what happens in $\AdS_5\times\S^5$~\cite{Arutyunov:2006ak, Arutyunov:2006yd}, we cannot use the coset description, since after gauge fixing massless fermions do not have a conventional quadratic kinetic term. Instead we will work with the Green-Schwarz action~\cite{Grisaru:1985fv}. In this way we find the algebra~$\mathcal{A}$ of ``off-shell'' symmetries---\textit{i.e.}, the ones of~$\Smat$---and use it to obtain the all-loop kinematics and S~matrix. We see in particular that massless modes are non-relativistic, with a giant-magnon dispersion relation~\cite{Hofman:2006xt}.
This facilitates our treatment and distinguishes it from the relativistic 
case~\cite{Zamolodchikov:1992zr}---in particular, no \textit{ad hoc} scaling limit needs to be taken. The more technical details of our analysis will be presented elsewhere~\cite{future:details}.

\section{Off-shell symmetry algebra}
 
To find the algebra~$\mathcal{A}$ we decompactify the worldsheet and take the theory off shell by relaxing the level matching condition.
The superisometries of the $\AdS_3 \times \S^3 \times \T^4$ string theory background form the algebra $\psu(1,1|2)_{\sL} \times \psu(1,1|2)_{\sR}$~\footnote{The $\u(1)^4$ shift isometries along the $\T^4$ directions will play no role in this letter.}, where the labels L and R refer to the left- and right-moving symmetries of the dual $\CFT_2$.
Only those charges that commute with the gauge-fixed Hamiltonian $\gen{H}$ sit in~$\mathcal{A}$. This gives eight supercharges and four central elements $\gen{H}$, $\gen{M}$, $\gen{C}$ and $\overline{\gen{C}}$, as well as an $\so(4) = \su(2)_{\suA} \times \su(2)_{\suB}$ algebra which arises in the decompactified theory because the winding modes on the torus decouple. The charge $\gen{M}$ is a combination of angular momenta on $\AdS_3$ and $\S^3$. 
The relaxed level matching condition yields the additional central charges $\gen{C}$ and $\overline{\gen{C}}$, which are not part of the isometry algebra and hence must vanish for physical on-shell states. 
At vanishing winding such states satisfy $\gen{P}\ket{\text{phys}}=0$, where $\gen{P}$ is the worldsheet momentum operator.

To proceed we need to gauge-fix the theory. We first impose the lightcone $\kappa$-gauge 
\begin{equation}
(\Gamma^0+\Gamma^5)\,\theta^I=0\,, \qquad I=1,2\,,
\label{kappagauge}
\end{equation}
where $X^0 = t$ and $X^5 = \phi$ parametrise the global time in $\AdS_3$ and a great circle along $\S^3$, respectively, $\theta^I$ are $9+1$-dimensional Majorana-Weyl fermions of Type IIB string theory and $\Gamma$ are $32\times 32$ Dirac gamma matrices. This $\kappa$-gauge guarantees a conventional kinetic term for the fermions, but is not compatible with the coset action~\cite{Babichenko:2009dk}. We therefore have to work with the superspace action~\cite{Grisaru:1985fv,Cvetic:1999zs,Wulff:2013kga}. 
To determine the structure of $\mathcal{A}$ it is sufficient to consider terms at most quadratic in 
the fermions, for which explicit expressions are given in Ref.~\cite{Cvetic:1999zs}.

Before $\kappa$-gauge fixing, the string theory action on $\AdS_3\times \S^3\times \T^4$ is invariant under constant shifts of 16 (suitably defined) fermions. The supercharges can be found using the Noether procedure. Upon fixing the lightcone $\kappa$-gauge, half of the shifts breaks the gauge-fixing condition~\eqref{kappagauge} and need to be combined with a compensating $\kappa$-transformation, just like in flat space~\cite{Green:1983wt}. The eight supercharges in $\mathcal{A}$ are of this kind.

To fix the bosonic gauge symmetry we work in the first-order formalism. Introducing the lightcone coordinates $X^\pm=\tfrac{1}{2}(\phi\pm t)$ and the conjugate momenta $P_\pm$ we impose the uniform lightcone gauge~\cite{Arutyunov:2004yx, Arutyunov:2009ga}
\begin{equation}
  X^+ = \tau , \qquad 
  P_- = \text{const.} ,
\end{equation}
where $\tau$ is the worldsheet time. The Virasoro constraints are used to determine $X^-$ and $P_+$ as non-local expressions in terms of the physical transverse fields. They are related to $\gen{H}$ and $\gen{P}$ by
\begin{equation}
\gen{H} = - \! \int_{-\infty}^{+\infty} \! \de\sigma \, P_+ , \qquad
\gen{P} = 2 \! \int_{-\infty}^{+\infty} \!  \de\sigma \, \partial_\sigma X^- .
\end{equation}
To carry out this procedure explicitly we need to redefine the fermions $\theta^I$ to make them  neutral under the $U(1)$ iso\-me\-tries generated by $t$ and $\phi$ translations~\cite{Alday:2005ww}. This yields a non-local piece in the supercharges. The supercharges' dependence on the massless fields $X^{\mu}$, on the conjugate momenta $P_{\mu}$, and on the massless fermions $\chi_I$ is
\begin{equation}\label{eq:supercharges}
  \begin{aligned}
    \gen{Q}_{\mu} &= \int_{-\infty}^{+\infty} \de\sigma \, e^{+\Gamma^{34}X^-}
    \left( P_{\mu} \chi_1 - \partial_\sigma X_{\mu} \chi_2 \right) ,
    \\
    \widetilde{\gen{Q}}_{\mu} & =\int_{-\infty}^{+\infty} \de\sigma \, e^{-\Gamma^{34}X^-}
    \left( P_{\mu} \chi_2 - \partial_\sigma X_{\mu} \chi_1 \right) ,
  \end{aligned}
\end{equation}
where $\mu=6,7,8,9$ runs over the $\T^4$ directions. This gives us eight real supercharges. We can combine them into two $\su(2)_{\suA}$ doublets $\gen{Q}_{\sL}^{\ a}$ and $\gen{Q}_{\sR a}$ and their complex conjugates $\overline{\gen{Q}}_{\sL a} = ( \gen{Q}_{\sL}^{\ a} )^\dag$ and $\overline{\gen{Q}}{}_{\sR}^{\ a} = ( \gen{Q}_{\sR a} )^\dag$. 
The L and R labels indicate which $\psu(1,1|2)$ superisometry algebra the charges originate from.
The supercharges satisfy the $\psu(1|1)^4_{\ce}$ algebra
\begin{equation}
\label{eq:commrel:su11:4}
  \begin{aligned}
    \acomm{\gen{Q}_{\sL a}}{\overline{\gen{Q}}{}_{\sL}^{\ b}} &= \tfrac{1}{2} \delta^b_a ( \gen{H} + \gen{M} ) , &\ \ 
    \acomm{\gen{Q}_{\sL a}}{\gen{Q}_{\sR}^{\ b}} &= \delta^b_a \gen{C} , \\
    \acomm{\gen{Q}_{\sR}^{\ a}}{\overline{\gen{Q}}_{\sR b}} &= \tfrac{1}{2} \delta^a_b ( \gen{H} - \gen{M} ) , &
    \acomm{\overline{\gen{Q}}{}_{\sL}^{\ a}}{\overline{\gen{Q}}_{\sR b}} &= \delta^a_b \overline{\gen{C}} ,
  \end{aligned}
\end{equation}
The central charge $\gen{C}$ and its conjugate $\overline{\gen{C}}$ arise from the non-local exponential factor in~\eqref{eq:supercharges} and are related to the worldsheet momentum by
\begin{equation}
\label{eq:offshellcharges}
  \gen{C} = \frac{i h \zeta}{2} ( e^{i \gen{P}} - 1 )\, ,
\end{equation}
where the phase $\zeta = e^{2iX^-(-\infty)}$ depends on the boundary conditions for the field $X^-$~\cite{Arutyunov:2006yd} and $h$ is the string tension. For a single excitation,~$\zeta$ can be absorbed by a rescaling, but for multi-excitation states the relative phases are relevant, as we will see later.

\section{Representations}

\begin{figure}[t]
  \centering
  \begin{tikzpicture}[%
    box/.style={outer sep=1pt},
    Q node/.style={inner sep=1pt,outer sep=0pt},
    arrow/.style={-latex}
    ]%

    \node [box] (PhiM) at ( 0  , 1.2cm) {\small $\ket{Y^{\sL}}$};
    \node [box] (PsiP) at (-1.2cm, 0cm) {\small $\ket{\eta^{\sL 1}}$};
    \node [box] (PsiM) at (+1.2cm, 0cm) {\small $\ket{\eta^{\sL 2}}$};
    \node [box] (PhiP) at ( 0  ,-1.2cm) {\small $\ket{Z^{\sL}}$};

    \newcommand{\horshift}{0.09cm,0cm}
    \newcommand{\vershift}{0cm,0.10cm}
 
    \draw [arrow] ($(PhiM.west) +(\vershift)$) -- ($(PsiP.north)-(\horshift)$) node [pos=0.5,anchor=south east,Q node] {\scriptsize $\gen{Q}^{\ 1}_{\sL}$};
    \draw [arrow] ($(PsiP.north)+(\horshift)$) -- ($(PhiM.west) -(\vershift)$) node [pos=0.5,anchor=north west,Q node] {};

    \draw [arrow] ($(PsiM.south)-(\horshift)$) -- ($(PhiP.east) +(\vershift)$) node [pos=0.5,anchor=south east,Q node] {};
    \draw [arrow] ($(PhiP.east) -(\vershift)$) -- ($(PsiM.south)+(\horshift)$) node [pos=0.5,anchor=north west,Q node] {\scriptsize $\overline{\gen{Q}}{}_{\sL 1}$};

    \draw [arrow] ($(PhiM.east) -(\vershift)$) -- ($(PsiM.north)-(\horshift)$) node [pos=0.5,anchor=north east,Q node] {};
    \draw [arrow] ($(PsiM.north)+(\horshift)$) -- ($(PhiM.east) +(\vershift)$) node [pos=0.5,anchor=south west,Q node] {\scriptsize $\overline{\gen{Q}}{}_{\sL 2}$};

    \draw [arrow] ($(PsiP.south)-(\horshift)$) -- ($(PhiP.west) -(\vershift)$) node [pos=0.5,anchor=north east,Q node] {\scriptsize $-\gen{Q}^{\ 2}_{\sL}$};
    \draw [arrow] ($(PhiP.west) +(\vershift)$) -- ($(PsiP.south)+(\horshift)$) node [pos=0.5,anchor=south west,Q node] {};

    \draw [arrow] (PsiM) -- (PsiP) node [pos=0.65,anchor=south west,Q node] {\scriptsize $\gen{J}^{\ a}_{\suA}$};
    \draw [arrow] (PsiP) -- (PsiM);

\draw[rounded corners=5mm] (-1.7cm,-1.5cm)rectangle (1.6cm,1.5cm);
  \end{tikzpicture}
\hspace{1cm}
  \begin{tikzpicture}[%
    box/.style={outer sep=1pt},
    Q node/.style={inner sep=1pt,outer sep=0pt},
    arrow/.style={-latex}
    ]%

    \node [box] (PhiM) at ( 0  , 1.2cm) {\small $\ket{Z^{\sR}}$};
    \node [box] (PsiP) at (-1.2cm, 0cm) {\small $\ket{\eta^{\sR}_{\  1}}$};
    \node [box] (PsiM) at (+1.2cm, 0cm) {\small $\ket{\eta^{\sR}_{\  2}}$};
    \node [box] (PhiP) at ( 0  ,-1.2cm) {\small $\ket{Y^{\sR}}$};

    \newcommand{\horshift}{0.09cm,0cm}
    \newcommand{\vershift}{0cm,0.10cm}
 
    \draw [arrow] ($(PsiP.north)-(\horshift)$) -- ($(PhiM.west) +(\vershift)$) node [pos=0.5,anchor=south east,Q node] {\scriptsize $\ -\gen{Q}_{\sR 2}$};
    \draw [arrow] ($(PhiM.west) -(\vershift)$) -- ($(PsiP.north)+(\horshift)$) node [pos=0.5,anchor=north west,Q node] {};

    \draw [arrow] ($(PhiP.east) +(\vershift)$) -- ($(PsiM.south)-(\horshift)$) node [pos=0.5,anchor=south east,Q node] {};
    \draw [arrow] ($(PsiM.south)+(\horshift)$) -- ($(PhiP.east) -(\vershift)$) node [pos=0.5,anchor=north west,Q node] {\scriptsize $\overline{\gen{Q}}{}^{\ 2}_{\sR}$};

    \draw [arrow] ($(PsiM.north)-(\horshift)$) -- ($(PhiM.east) -(\vershift)$) node [pos=0.5,anchor=north east,Q node] {};
    \draw [arrow] ($(PhiM.east) +(\vershift)$) -- ($(PsiM.north)+(\horshift)$) node [pos=0.5,anchor=south west,Q node] {\scriptsize $\overline{\gen{Q}}{}^{\ 1}_{\sR}$};

    \draw [arrow] ($(PhiP.west) -(\vershift)$) -- ($(PsiP.south)-(\horshift)$) node [pos=0.5,anchor=north east,Q node] {\scriptsize $\gen{Q}_{\sR 1}$};
    \draw [arrow] ($(PsiP.south)+(\horshift)$) -- ($(PhiP.west) +(\vershift)$) node [pos=0.5,anchor=south west,Q node] {};

    \draw [arrow] (PsiM) -- (PsiP) node [pos=0.65,anchor=south west,Q node] {\scriptsize $\gen{J}^{\ a}_{\suA}$};
    \draw [arrow] (PsiP) -- (PsiM);
    
\draw[rounded corners=5mm] (-1.7cm,-1.5cm)rectangle (1.6cm,1.5cm);
  \end{tikzpicture}
  \caption{Each of the two (left and right) massive $\psu(1|1)^4_{\ce}$ multiplets consists of two bosons~$Y^{\sL,\sR}$, $Z^{\sL,\sR}$ and of two fermions~$\eta_{\ a}^{\sL,\sR}$, the latter carrying the fundamental $\su(2)_{\suA}$ index~$a$. For clarity we only indicate the supercharges that do not vanish on~shell.}
  \label{fig:massive}
\end{figure}
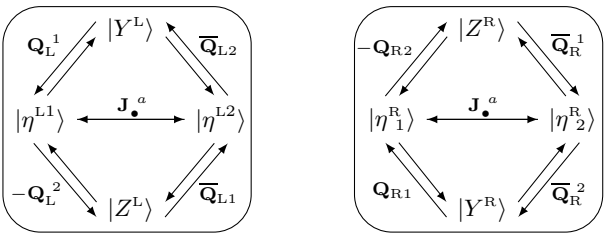

The fundamental excitations of the string---eight bosons and eight fermions---arrange themselves into \emph{short} multiplets of the off-shell symmetry algebra~$\mathcal{A}$, satisfying the shortening condition \begin{equation}
\label{eq:shortening}
\gen{H}^2=\gen{M}^2+4\,\overline{\gen{C}}\,\gen{C}\,.
\end{equation}
Since $\gen{C}$ vanishes at zero momentum, the eigenvalue~$m$ of~$\gen{M}$ plays the role of a mass. There are two bosonic and two fermionic excitations with mass $m=+1$, and we refer to them as \emph{left-flavoured} because on-shell they transform only under the left supercharges. Similarly, there are four \emph{right-flavoured} excitations with $m=-1$, and eight \emph{massless} excitations with $m=0$.
The corresponding multiplets are depicted in figures~\ref{fig:massive} and~\ref{fig:massless}. There we see four $\psu(1|1)^4_{\ce}$ \emph{bifundamental} representations, supplemented by the action of $\su(2)_{\suA}$ and $\su(2)_{\suB}$, the latter acting on massless excitations only.

The algebra~\eqref{eq:commrel:su11:4} can be obtained from two copies of $\psu(1|1)^2_{\ce}$. This consists of four conjugate supercharges $\bar{\sgen{q}}_{\sL,\sR}=(\sgen{q}_{\sL,\sR})^{\dagger}$, satisfying
\begin{equation}
\begin{aligned}
\left\{\sgen{q}_{\sL},\bar{\sgen{q}}_{\sL}\right\}&=\tfrac{1}{2}\big(\sgen{h}+\sgen{m}\big),
\quad &
\left\{\sgen{q}_{\sL},{\sgen{q}}_{\sR}\right\}&=\sgen{c}\,,\\
\left\{\sgen{q}_{\sR},\bar{\sgen{q}}_{\sR}\right\}&=\tfrac{1}{2}\big(\sgen{h}-\sgen{m}\big),
\quad&
\left\{\bar{\sgen{q}}_{\sL},\bar{\sgen{q}}_{\sR}\right\}&=\bar{\sgen{c}}\,.
\end{aligned}
\end{equation}
We can then set~$\gen{Q}^{\ 1}_{\sL}=\sgen{q}_{\sL}\otimes\id$,
$\gen{Q}^{\ 2}_{\sL}=\id\otimes\sgen{q}_{\sL}$ for the left~flavour, $\gen{Q}_{\sR 1}=\sgen{q}_{\sR}\otimes\id$,
$\gen{Q}_{\sR 2}=\id\otimes\sgen{q}_{\sR}$ for the right, and similarly for their conjugates.
The bifundamental representations of~$\psu(1|1)^4_{\ce}$ can be obtained from the \emph{fundamental} representations of~$\psu(1|1)^2_{\ce}$. One such representation, which we denote by~$\varrho_{\sL}=(\phi^{\sL}|\psi^{\sL})$, is
\begin{equation}
\begin{aligned}
\sgen{q}_{\sL}\ket{\phi_{p}^{\sL}}&=a_p\ket{\psi_{p}^{\sL}},
&&\qquad
\bar{\sgen{q}}_{\sL}\ket{\psi_{p}^{\sL}}=\bar{a}_p\ket{\phi_{p}^{\sL}},\\
\sgen{q}_{\sR}\ket{\psi_{p}^{\sL}}&=b_p\ket{\phi_{p}^{\sL}},
&&\qquad
\bar{\sgen{q}}_{\sR}\ket{\phi_{p}^{\sL}}=\bar{b}_p\ket{\psi_{p}^{\sL}},
\end{aligned}
\end{equation}
where the representation coefficients depend on the excitation momentum~$p$.  
Another representation,~$\varrho_{\sR}$ can be obtained by exchanging the action of left and right generators. Two more representations~$\widetilde{\varrho}_{\sL}$, $\widetilde{\varrho}_{\sR}$, can be obtained by exchanging bosons and fermions.

Using this, the left and right representations of figure~\ref{fig:massive} are given by~$\varrho_{\sL}\otimes\varrho_{\sL}$ and~$\varrho_{\sR}\otimes\varrho_{\sR}$ respectively. On the former, the central charges are
\begin{equation}
\begin{aligned}
\gen{H}&=\big(a_p\,\bar{a}_p+b_p\,\bar{b}_p\big)\id\,,
&&\qquad
\gen{C}=a_p\,b_p\,\id\,,\\
\gen{M}&=\big(a_p\,\bar{a}_p-b_p\,\bar{b}_p\big)\id\,,
&&\qquad
\overline{\gen{C}}=\bar{a}_p\,\bar{b}_p\,\id\,,
\end{aligned}
\end{equation}
while on the latter one should exchange~$a_p \leftrightarrow b_p$, flipping the sign of~$\gen{M}$.
We then see that whole massive module is invariant under relabelling~$\bL\leftrightarrow\bR$, resulting in a~$\mathbb{Z}_2$ \emph{left-right} (LR) symmetry~\cite{Borsato:2012ud,Borsato:2013qpa}. 

The two massless~$\psu(1|1)^4_{\ce}$ modules have a fermionic highest weight, and up to a change of basis they can equivalently be given by $\varrho_{\sL}\otimes\widetilde{\varrho}_{\sL}$ or $\varrho_{\sR}\otimes\widetilde{\varrho}_{\sR}$~\footnote{%
Taking the same form for both~$\psu(1|1)^4_{\ce}$ modules makes the~$\su(2)_{\suB}$ invariance more manifest, while having two different forms makes crossing symmetry and the completion of LR~symmetry in the massless sector more explicit.%
}, provided that all representation parameters satisfy $a_p\,\bar{a}_p=b_p\,\bar{b}_p$,
\textit{i.e.} that $\gen{M}$ vanishes. This is not only a semiclassical input, but a consistency condition:~$\su(2)_{\suB}$ invariance requires $\gen{M}$ to take the same value on both modules, while crossing invariance requires~$\gen{M}$ to take opposite values.

\begin{figure}[t]
  \centering
  \begin{tikzpicture}[%
    box/.style={outer sep=1pt},
    Q node/.style={inner sep=1pt,outer sep=0pt},
    arrow/.style={-latex}
    ]%
\begin{scope}[xshift=-2.4cm]
    \node [box] (PhiM) at ( 0  , 1.2cm) {\small $\ket{\chi^{1}}$};
    \node [box] (PsiP) at (-1.2cm, 0cm) {\small $\ket{T^{11}}$};
    \node [box] (PsiM) at (+1.2cm, 0cm) {\small $\ket{T^{21}}$};
    \node [box] (PhiP) at ( 0  ,-1.2cm) {\small $\ket{\widetilde{\chi}^{1}}$};

    \newcommand{\horshift}{0.09cm,0cm}
    \newcommand{\vershift}{0cm,0.10cm}
 
    \draw [arrow] ($(PhiM.west) +(\vershift)$) -- ($(PsiP.north)-(\horshift)$) node [pos=0.5,anchor=south east,Q node] {\scriptsize $\gen{Q}^{\ 1}_{\sL}$};
    \draw [arrow] ($(PsiP.north)+(\horshift)$) -- ($(PhiM.west) -(\vershift)$) node [pos=0.5,anchor=north west,Q node] {};

    \draw [arrow] ($(PsiM.south)-(\horshift)$) -- ($(PhiP.east) +(\vershift)$) node [pos=0.5,anchor=south east,Q node] {};
    \draw [arrow] ($(PhiP.east) -(\vershift)$) -- ($(PsiM.south)+(\horshift)$) node [pos=0.5,anchor=north west,Q node] {\scriptsize $\overline{\gen{Q}}{}_{\sL 1}$};

    \draw [arrow] ($(PhiM.east) -(\vershift)$) -- ($(PsiM.north)-(\horshift)$) node [pos=0.5,anchor=north east,Q node] {};
    \draw [arrow] ($(PsiM.north)+(\horshift)$) -- ($(PhiM.east) +(\vershift)$) node [pos=0.5,anchor=south west,Q node] {\scriptsize $\overline{\gen{Q}}{}_{\sL 2}$};

    \draw [arrow] ($(PsiP.south)-(\horshift)$) -- ($(PhiP.west) -(\vershift)$) node [pos=0.5,anchor=north east,Q node] {\scriptsize $-\gen{Q}^{\ 2}_{\sL}$};
    \draw [arrow] ($(PhiP.west) +(\vershift)$) -- ($(PsiP.south)+(\horshift)$) node [pos=0.5,anchor=south west,Q node] {};

    \draw [arrow] (PsiM) -- (PsiP) node [pos=0.65,anchor=south west,Q node] {\scriptsize $\gen{J}^{\ a}_{\suA}$};
    \draw [arrow] (PsiP) -- (PsiM);

\draw[rounded corners=5mm] (-1.7cm,-1.5cm)rectangle (1.6cm,1.5cm);
\end{scope}
\begin{scope}[xshift=0cm]
    \draw [arrow] (-0.6cm,0cm) -- (0.6cm,0cm) node [Q node] at (0cm,0.15cm) {\scriptsize $\gen{J}^{\ \alpha}_{\suB}$};
    \draw [arrow] (0.6cm,0cm) -- (-0.6cm,0cm);
    %
\end{scope}
\begin{scope}[xshift=2.5cm]

    \node [box] (PhiM) at ( 0  , 1.2cm) {\small $\ket{\chi^{2}}$};
    \node [box] (PsiP) at (-1.2cm, 0cm) {\small $\ket{T^{12}}$};
    \node [box] (PsiM) at (+1.2cm, 0cm) {\small $\ket{T^{22}}$};
    \node [box] (PhiP) at ( 0  ,-1.2cm) {\small $\ket{\widetilde{\chi}^{2}}$};

    \newcommand{\horshift}{0.09cm,0cm}
    \newcommand{\vershift}{0cm,0.10cm}
 
    \draw [arrow] ($(PhiM.west) +(\vershift)$) -- ($(PsiP.north)-(\horshift)$) node [pos=0.5,anchor=south east,Q node] {\scriptsize $\gen{Q}^{\ 1}_{\sL}$};
    \draw [arrow] ($(PsiP.north)+(\horshift)$) -- ($(PhiM.west) -(\vershift)$) node [pos=0.5,anchor=north west,Q node] {};

    \draw [arrow] ($(PsiM.south)-(\horshift)$) -- ($(PhiP.east) +(\vershift)$) node [pos=0.5,anchor=south east,Q node] {};
    \draw [arrow] ($(PhiP.east) -(\vershift)$) -- ($(PsiM.south)+(\horshift)$) node [pos=0.5,anchor=north west,Q node] {\scriptsize $\overline{\gen{Q}}{}_{\sL 1}$};

    \draw [arrow] ($(PhiM.east) -(\vershift)$) -- ($(PsiM.north)-(\horshift)$) node [pos=0.5,anchor=north east,Q node] {};
    \draw [arrow] ($(PsiM.north)+(\horshift)$) -- ($(PhiM.east) +(\vershift)$) node [pos=0.5,anchor=south west,Q node] {\scriptsize $\overline{\gen{Q}}{}_{\sL 2}$};

    \draw [arrow] ($(PsiP.south)-(\horshift)$) -- ($(PhiP.west) -(\vershift)$) node [pos=0.5,anchor=north east,Q node] {\scriptsize $-\gen{Q}^{\ 2}_{\sL}$};
    \draw [arrow] ($(PhiP.west) +(\vershift)$) -- ($(PsiP.south)+(\horshift)$) node [pos=0.5,anchor=south west,Q node] {};

    \draw [arrow] (PsiM) -- (PsiP) node [pos=0.65,anchor=south west,Q node] {\scriptsize $\gen{J}^{\ a}_{\suA}$};
    \draw [arrow] (PsiP) -- (PsiM);

\draw[rounded corners=5mm] (-1.7cm,-1.5cm)rectangle (1.6cm,1.5cm);
\end{scope}
  \end{tikzpicture}
  \caption{The two massless $\psu(1|1)^4_{\ce}$ multiplets, in the representation $(\varrho_{\sL}\otimes\widetilde{\varrho}_{\sL})^{\oplus2}$. Overall we have four bosons $T^{a\alpha}$~and four fermions~$\chi^{\alpha}, \widetilde{\chi}^{\alpha}$, where $a$ and $\alpha$ are fundamental indices of $\su(2)_{\suA}$ and $\su(2)_{\suB}$. Again we show only some supercharges. Note that~$\su(2)_{\suB}$ relates the two $\psu(1|1)^4_{\ce}$~modules.}
  \label{fig:massless}
\end{figure}

The explicit form~\eqref{eq:offshellcharges} of~$\gen{C},\overline{\gen{C}}$ and the shortening condition~\eqref{eq:shortening} yield the dispersion relation~\cite{David:2008yk,David:2010yg}
\begin{equation}
\label{eq:dispersion}
E_p=\sqrt{m^2+4h^2\sin^2 \frac{p}{2} }\,. 
\end{equation}
In particular, for massless excitations the dispersion~$E_p=2h\,|\sin(p/2)|$ is non-analytic. This can be resolved by treating worldsheet left- and right-movers separately, as typical for massless two-dimensional excitations.
One may worry that masslessness is spoiled by quantum corrections, as in other integrable models, \textit{e.g.}~\cite{Gross:1974jv}. This is impossible unless some symmetry is broken, since a dynamical mass would correct the eigenvalue of~$\gen{M}$.

We can construct the two-particle representations, on which the S~matrix acts, out of pairs of one-particle ones. This introduces a non-local momentum dependence through the phase $\zeta$ \cite{Arutyunov:2006ak,Arutyunov:2006yd}. In a Hopf-algebra language, this amounts to defining a deformed coproduct~\cite{Borsato:2013qpa,Arutyunov:2006yd}.

\section{All-loop S~matrix}
The two-body S~matrix~$\Smat (p,q)$ must be invariant under~$\mathcal{A}$. Furthermore, it must satisfy braiding and physical unitarity, crossing symmetry, and the aforementioned Yang-Baxter equation
~\cite{Arutyunov:2009ga,future:review}. 
Here these conditions are restrictive enough to fix~$\Smat$ up to few scalar factors denoted by~$\sigma$, which must obey non-trivial constraints. 

Scattering processes involving massless particles may appear ill-defined. In relativistic theories indeed massless wave-packets travel at the speed of light, and in $1+1$~dimension na\"ively cannot scatter if they move in the same direction~\cite{Zamolodchikov:1992zr}. Here, instead, the non-relativistic dispersion~\eqref{eq:dispersion} at zero mass yields the group velocity $\partial E_p /\partial p = \pm 2 h \cos (p/2)$. Therefore, massless wave-packets with different momenta generically scatter.

The two-body S~matrix naturally decomposes into the \emph{massive} ($\bullet\bullet$), \emph{mixed} ($\bullet\circ, \circ\bullet$) and \emph{massless} ($\circ\circ$) sectors, depending on the masses of the excitations scattering:
\begin{equation}
\Smat = \!\left(\!
      \begin{array}{cc}
        \Smat^{\bullet\bullet} & \Smat^{\circ\bullet}  \\
        \Smat^{\bullet\circ} & \Smat^{\circ\circ}  \\
      \end{array}\!\right).
\end{equation}
In each sector it further breaks into several blocks, each scattering $\psu(1|1)^4_{\ce}$ irreducible representations.
Some of the scalar factors multiplying each block are then related by LR or $\su(2)_\circ$ symmetry. Exploiting the bifundamental nature of the representations, we write the blocks as graded tensor products of $\su(1|1)^2_{\ce}$-invariant S~matrices. These were computed in Ref.~\cite{Borsato:2012ud} for the representations $\varrho_{\sL}, \varrho_{\sR}$ at any value of the mass. We denote them by $\Smat^{\sL\sL},$ $\Smat^{\sR\sR},\Smat^{\sR\sL},\Smat^{\sL\sR}$.
The remaining S~matrices involving $\varrho_{\stL}$ are obtained from these by a similarity transformation exchanging the boson and the fermion, yielding \emph{e.g.}~$\Smat^{\stL\sL}$.

\paragraph{Massive sector.} We have four blocks
\begin{equation}
  \Smat^{\bullet\bullet} = \!\left(\!
    \begin{array}{ccc}
      \sigma^{\bullet\bullet}\; \Smat^{\sL\sL} \hat{\otimes} \Smat^{\sL\sL} & & \widetilde{\sigma}^{\bullet\bullet}\; \Smat^{\sR\sL} \hat{\otimes} \Smat^{\sR\sL} \\[4pt]
      \widetilde{\sigma}^{\bullet\bullet}\; \Smat^{\sL\sR} \hat{\otimes} \Smat^{\sL\sR} & & \sigma^{\bullet\bullet}\; \Smat^{\sR\sR} \hat{\otimes} \Smat^{\sR\sR}
    \end{array}\!\right), 
\end{equation}
where $\hat{\otimes}$ denotes the graded tensor product.
On the diagonal we have left-left and right-right scattering. Owing to LR symmetry the corresponding S~matrices coincide.
On the anti-diagonal we have opposite-flavour S~matrices, also related by LR symmetry.
There are then just two independent scalar factors in the massive sector: $\sigma^{\bullet\bullet}$ and~$\widetilde{\sigma}^{\bullet\bullet}$. This agrees with Ref.~\cite{Borsato:2013qpa}, where the massive sector was studied in a spin-chain framework~\footnote{In particular our scalar factors ${\sigma^{\bullet\bullet}}, {\protect\widetilde{\sigma}^{\bullet\bullet}}$ are directly related to ${\sigma}, {\protect\widetilde{\sigma}}$ there.
}.

\paragraph{Mixed sector.} This involves massive-massless and massless-massive  scattering. In the former case we find
\begin{equation}
\Smat^{\bullet\circ} = \sigma^{\bullet\circ} \; \left[ \left( \Smat^{\sL\sL} \hat{\otimes} \Smat^{\sL\stL} \right) \oplus \left( \Smat^{\sR\sL}  \hat{\otimes} \Smat^{\sR\stL} \right) \right]^{\oplus 2}.
\end{equation}
The direct sum inside the square brackets corresponds to scattering with either of two $\psu(1|1)^4_\ce$ massive modules---left or  right. These two S~matrices are then identified after imposing LR symmetry, which is possible because the second excitation is massless. Since scattering can occur with two different $\psu(1|1)^4_{\ce}$ massless modules we have two copies of the expression inside the square brackets.
These must be equal by  $\su(2)_{\circ}$ symmetry. Owing to these symmetries we are left with a single undetermined scalar factor~$\sigma^{\bullet\circ}$.
Similar considerations apply to massless-massive scattering, yielding
\begin{equation}
\Smat^{\circ\bullet} = \sigma^{\circ\bullet} \; \left[ \left( \Smat^{\sL\sL} \hat{\otimes} \Smat^{\stL\sL} \right) \oplus \left( \Smat^{\sL\sR} \hat{\otimes} \Smat^{\stL\sR} \right) \right]^{\oplus 2},
\end{equation} 
where we have another scalar factor $\sigma^{\circ\bullet}$. 

\paragraph{Massless sector.}  Here the S~matrix factorises as 
\begin{equation}
\Smat^{\circ\circ} =\sigma^{\circ\circ} \; \Smat_{\su(2)} \otimes \big(\Smat^{\sL\sL} \hat{\otimes} \Smat^{\stL\stL}\big),
\end{equation}
where the factors in brackets are fixed by the $\psu(1|1)^4_{\ce}$ invariance. We have a single scalar factor $\sigma^{\circ\circ}$ and an $\su(2)_{\circ}$-invariant S~matrix
\begin{equation}
\Smat_{\su(2)} (p,q) = \frac{1}{1+ i(w_p-w_q) } \left(\Pi + i (w_p-w_q) \mathbf{1}_4 \right),
\end{equation}
with $\Pi$ the permutation operator and $w_p$ a real function of the momentum $p$. This is the Heisenberg-model S~matrix, where $w_p$ plays the role of a generalised rapidity.

\paragraph{Unitarity.}
The S~matrix satisfies braiding and physical unitarity, which result in constraints on the scalar factors.
These are solved by taking~$\sigma^{\bullet\bullet}, \widetilde{\sigma}^{\bullet\bullet}$ and~$\sigma^{\circ\circ}$ to be exponentials of anti-symmetric phases in a suitable normalisation, and by simply relating $\sigma^{\bullet\circ}$ to $\sigma^{\circ\bullet}$. 

\paragraph{Crossing symmetry.}
The crossing transformation maps a representation to its conjugate, flipping the sign of all central charges including momentum and energy. This requires analytic continuation to an unphysical channel~\cite{Arutyunov:2009ga, future:review, Janik:2006dc}.
It is defined through the charge conjugation matrix~$\Ccr$, which decomposes on the representations,~$\Ccr = \Ccr^\bullet \oplus \Ccr^\circ$. The massive-sector block~$\Ccr^\bullet$ was given in Ref.~\cite{Borsato:2013qpa}, and yields~$\Ccr^\circ$ by a similarity transformation and by requiring compatibility with~$\su(2)_{\suB}$~\cite{future:details}.
Such transformation is momentum-dependent, so that~$\Ccr_p^{\circ}$ depends discontinuously on the worldsheet chirality through $\operatorname{sign}\left(\sin (p/2)\right)$---a signature of the massless modes.
Crossing invariance of~$\Smat$ requires
\begin{equation}\label{eq:cr-matrix-form}
(\mathbf{1}  \otimes \Ccr_q^{-1}) \cdot \Smat^{\text{t}_2}(p,\bar{q}) \cdot (\mathbf{1} \otimes \Ccr_{q}^{\phantom{1}}) \cdot \Smat(p,q) = \mathbf{1}\otimes \mathbf{1},
\end{equation}
with $\text{t}_2$ meaning transposition in the second space and~$\bar{q}$ analytic continuation. 
This matrix equation results in five equations for the scalar factors. Two of them constrain $\sigma^{\bullet\bullet}$ and~$ \widetilde{\sigma}^{\bullet\bullet}$, and were solved in Ref.~\cite{Borsato:2013hoa}. Each of the other three equations involves one of the remaining scalar factors $\sigma^{\bullet\circ}, \sigma^{\circ\bullet}$ and $\sigma^{\circ\circ}$.
Crossing also constrains the function $w_{p}$, setting $w_{\bar{p}}=w_p-i$.

\section{Outlook}

In this letter we have determined, up to a number of phases, the complete all-loop S matrix for fundamental excitations of pure-RR $\AdS_3\times\S^3\times\T^4$ strings.
An immediate next step is solving the crossing equations for $\sigma^{\circ\bullet},\sigma^{\bullet\circ}$ and~$\sigma^{\circ\circ}$ as done for $\sigma^{\bullet\bullet}$ and~$\widetilde{\sigma}^{\bullet\bullet}$ in Ref.~\cite{Borsato:2013hoa}. This will likely present us with new analytic structures and may require further insights from perturbative calculations. It would be also interesting to write down the Bethe-Yang equations for the asymptotic spectrum and to see how the $\mathcal{N}=(4,4)$ symmetry is realised there. Finding the bound-state spectrum and S~matrix would then lead to the string hypothesis and mirror thermodynamical Bethe ansatz for the exact spectrum.
The success of the integrability approach on the~$\AdS_3$ side strongly suggests that an analogous description should exists for the~$\CFT_2$. It would be important to uncover it, perhaps building on Ref.~\cite{Pakman:2009mi}.

It should be possible to extend this approach to consider orbifolds and integrable deformations of this background---as it was successfully  done for $\AdS_5\times\S^5$. This might lead to new insights into $\AdS_3$ black holes~\cite{Banados:1992wn,Strominger:1996sh} and their integrability properties~\cite{David:2011iy}.
The methods presented in this letter can also be applied  to $\AdS_3\times\S^3\times\S^3\times\S^1$ superstrings~\cite{Borsato:2012ud,  Borsato:2012ss, Elitzur:1998mm}, whose dual CFT remains elusive~\cite{Gukov:2004ym,Tong:2014yna}. There, higher spin theories were recently considered~\cite{Gaberdiel:2013vva}, and integrability may help investigating their relation with strings.

Another interesting direction is to consider backgrounds with mixed RR and NSNS fluxes~\cite{Cagnazzo:2012se,Hoare:2013ida,Hoare:2013lja}. This may offer new insights on the relation between the infinitely-many conserved charges from integrability with the Virasoro ones appearing in the worldsheet CFT---perhaps along the lines of what happens in relativistic models~\cite{Fendley:1993jh}---as well as on how S~duality is implemented.

We are confident that there will be significant development in these directions in the near future.

\bigskip
\paragraph{Acknowledgements.}
We thank G.~Arutyunov, C.~Hull and K.~Zarembo for their careful reading of the manuscript.
R.B.\@ and A.S.\@ acknowledge support by the Netherlands Organization for Scientific Research (NWO) under the VICI grant 680-47-602. 
Their work is part of the ERC Advanced grant research programme No.\ 246974,
and of the D-ITP consortium, a program of the NWO that is funded by the Dutch Ministry of Education, Culture and Science (OCW).
O.O.S.'s  work was supported by the ERC Advanced grant No.\ 290456.
A.S.'s work is also funded by the European Union, Grant Agreement No.\ 317089. 
B.S.\@ acknowledges funding support from an STFC Consolidated Grant 
ST/J00037X/1.

\bibliographystyle{h-physrev}
\bibliography{massless}

\end{document}